\documentclass[11pt,a4paper]{article}
\pdfoutput=1
\usepackage{cancel}
\usepackage{enumitem}
\usepackage{etex}
\usepackage{amsthm}
\usepackage{geometry}
\usepackage{dcolumn}
\usepackage{epsf}
\usepackage{mathrsfs}
\usepackage{multirow}
\usepackage{booktabs}
\usepackage{tabularx}
\usepackage{array}
\usepackage{slashed}
\usepackage{float}
\usepackage{leftidx}
\usepackage{setspace}
\usepackage{verbatim}
\usepackage{adjustbox}
\usepackage{tikz}
\usepackage[caption=false]{subfig}
\usepackage{arydshln}
\usepackage{jheppub}
\usepackage{shuffle}
\usepackage{amsmath}
\usepackage{cases}
\usepackage{lipsum}
\usepackage{tabularx}
\usepackage{graphicx}
\usepackage{lscape}

\usepackage{lineno}
\usepackage{physics}
\usepackage{amssymb}
\usepackage{mathrsfs}
\usepackage{cancel} 

\numberwithin{equation}{section}
\usetikzlibrary{arrows,decorations.markings,shapes.arrows,patterns,positioning}

\newcommand{\bea}{\begin{eqnarray}}
\newcommand{\eea}{\end{eqnarray}}
\newcommand{\bean}{\begin{eqnarray*}}
\newcommand{\eean}{\end{eqnarray*}}
\newcommand{\nn}{\nonumber\\}
\newcommand{\Sl}{\sum\limits}

\def\W #1{\widetilde{#1}}

\def\Label#1{\label{#1}%
  \smash{\hbox to0pt{\raise1ex\hbox{\tiny[#1]}\hss}}}
\def\Label#1{\label{#1}}
\renewcommand{\eqref}[1]{eq.~(\ref{#1})}
\newcommand{\figref}[1]{Fig.~\ref{#1}}

\newcommand{\secref}[1]{section~\ref{#1}}

\def\braket#1{\left\langle #1 \right\rangle}

\def\vev{\braket}

\def\bvev#1{\left[ #1 \right]}
\def\Spaa{\vev}
\def\Spbb{\bvev}

\def\Sl{\sum\limits}

\newcommand{\ctobedelete}[1]{}

\allowdisplaybreaks

\DeclareFontFamily{U}{shuffle}{}
\DeclareFontShape{U}{shuffle}{m}{n}{ <-8>shuffle7 <8->shuffle10}{}

\title{\boldmath Note on single-trace EYM amplitudes with MHV configuration}

\author{Zhirun Li} \author{Yi-Jian Du}
\affiliation{School of Physics and Technology, Wuhan University,\\
No.299 Bayi Road, Wuhan 430072, P.R. China}

\emailAdd{zhirun.li@whu.edu.cn} \emailAdd{yijian.du@whu.edu.cn} 

\abstract{In the maximally-helicity-violating (MHV) configuration, tree-level single-trace Einstein-Yang-Mills (EYM) amplitude with one and two gravitons have been shown to satisfy a formula where each graviton splits into a pair of collinear gluons. In this paper, we extend this formula to more general cases. We provide a general formula which expresses tree-level single-trace MHV amplitudes in terms of pure gluon amplitudes, where each graviton turns into a pair of collinear gluons. }

\begin{document}
\maketitle
\flushbottom

\section{Introduction}
\label{sec:intro}

In four dimensional spacetime,  tree-level single-trace maximally-helicity-violating (MHV) amplitudes of Einstein-Yang-Mills (EYM) theory have been shown to satisfy the Selivanov-Bern-De Freitas-Wong \cite{Selivanov:1997aq, Selivanov:1997ts,Bern:1999bx} (SBDW) formula, which expresses the amplitude via a generating function. On another hand, Cachazo-He-Yuan (CHY) \cite{Cachazo:2013gna,Cachazo:2013hca,Cachazo:2013iea} formula gives a general approach to EYM amplitudes, which is independent of the dimension of spacetime and the helicity configuration. In four dimensions, the CHY formula has been shown to provide a {\it spanning forest formula} (first proposed in gravity, along the line of \cite{Nguyen:2009jk}, \cite{Hodges:2012ym} and \cite{Feng:2012sy}) for the single-trace MHV amplitude \cite{Du:2016wkt}, which was further proven to be equivalent with the SBDW formula \cite{Du:2016wkt} and was generalized to double-trace MHV amplitudes \cite{Tian:2021dzf} via the recursion expansion formula \cite{Fu:2017uzt,Chiodaroli:2017ngp,Teng:2017tbo,Du:2017kpo,Du:2017gnh}.

From another perspective, as pointed out in earlier literatures \cite{Stieberger:2009hq,Chen:2010sr,Stieberger:2014cea,Stieberger:2015qja,Stieberger:2015kia}, each graviton in an EYM amplitude could be considered as a pair of collinear gluons which carry the same momentum and the same helicity. Particularly, inspired by the SBDW formula, \cite{Chen:2010sr}  pointed out that the single-trace MHV amplitude with one and two gravitons can be explicitly expressed in terms of the MHV amplitudes where each graviton splits into a pair of collinear gluons \cite{Chen:2010sr}. This explicit formula of the single-trace MHV amplitudes was not extended into cases with an arbitrary number of gravitons yet. In this note, we take a small step forward in this direction: {\it we provide a general formula for single-trace MHV amplitudes where each graviton splits into a pair of collinear gluons.} When the number of gravitons is one or two, this formula turns back into the known results \cite{Chen:2010sr}. We hope this approach may provide a new insight for the study of helicity amplitudes in EYM.

The structure of this note is arranged as follows. In \secref{sec:Backgrounds}, a helpful review of spinor-helicity formalism and the SBDW formula is presented. We study the amplitude with three gravitons in \secref{sec:Example} and sketch the general proof in \secref{sec:GeneralFormula}. Further discussions and conclusions are presented in \secref{sec:Conclusions}.

\section{Backgrounds}\label{sec:Backgrounds}

In this section, we provide a brief review of the spinor-helicity formalism in four dimensions \cite{Xu:1986xb}, as well as the SBDW \cite{Selivanov:1997aq,Selivanov:1997ts,Bern:1999bx} formula and the spanning forest formula \cite{Du:2016wkt} for single-trace EYM amplitudes. 

\subsection{Spinor-helicity formalism in four dimensions}
The momentum $k^{\mu}_i$ of each on-shell massless particle $i$ is expressed by two copies of Weyl spinors $\lambda_i^{a}\W{\lambda}_i^{\dot{a}}$.  We define the spinor products as 
\bea
\Spaa{i,j}\equiv \epsilon_{a{b}}\lambda_i^{a}{\lambda}_{j}^{{b}},~~~~~~~~~~\Spbb{i,j}\equiv \epsilon_{\dot{a}\dot{b}}\W\lambda_i^{\dot{a}}\W{\lambda}_{j}^{\dot{b}}, \Label{Eq:SpinorProducts}
\eea
where $\epsilon_{{a}{b}}$ and $\epsilon_{\dot{a}\dot{b}}$ are totally antisymmetric tensors. Apparently, the spinor products are antisymmetric objects under the exchanging of the two spinors. With this expression, the Lorentz contraction of two momenta $k^{\mu}_a$ and $k^{\mu}_b$ reads:
\bea
k_a\cdot k_b=\frac{1}{2}\Spaa{a,b}\Spbb{b,a}. \Label{Eq:LorentzContractions}
\eea
More helpful properties in spinor-helicity formalsim are displayed as follows.
\begin{itemize}

\item Momentum conservation for an $n$-point amplitude:
\bea
\Sl_{\substack{i\neq \,j,k\\i=1}}^n\Spbb{j,i}\Spaa{i,k}=0.~~~~\Label{Eq:SpinorProp4}
\eea

\item Schouten identity:
\bea
\Spaa{a,b}\Spaa{c,d}=\Spaa{a,c}\Spaa{b,d}+\Spaa{b,c}\Spaa{d,a},~~~ \Spbb{a,b}\Spbb{c,d}=\Spbb{a,c}\Spbb{b,d}+\Spbb{b,c}\Spbb{d,a}.\Label{Eq:SpinorProp2}
\eea
\item The eikonal identity resulted by Schouten identity
\bea
\Sl_{i=j}^{k-1}\frac{\Spaa{i,i+1}}{\Spaa{i,q}\Spaa{q,i+1}}=\frac{\Spaa{j,k}}{\Spaa{j,q}\Spaa{q,k}}.~~~~~~~\Label{Eq:SpinorProp3}
\eea
\end{itemize}
Finally, the $n$-gluon MHV amplitude $A(1,...,n)$ at tree level satisfies the famous Parke-Taylor formula \cite{Parke:1986gb}\footnote{In this work, we use $\sim$ to neglect the coupling constant and an overall normalization factor.}:
\bea
A(1,...,n)\sim \frac{\langle ij\rangle^{4}}{\langle12\rangle\langle23\rangle\cdots\langle n1\rangle},
\eea
where $i$, $j$ denote the two negative-helicity gluons and other gluons are supposed to be positive-helicity ones.

\subsection{SBDW formula and the spanning forest formula}\label{sec:SBDWSpanning}
In EYM, there are two possible situations of the tree-level single-trace MHV amplitudes: the ($g^-,g^-$) and ($h^-,g^-$) configurations, which correspond to amplitudes with two negative-helicity gluons, and one negative-helicity gluon plus one negative-helicity graviton. In the following, we focus on the ($g^-,g^-$) configuration. The ($h^-,g^-$) can be studied similarly.

The SBDW \cite{Selivanov:1997aq,Selivanov:1997ts,Bern:1999bx} formula expresses the single trace ($g^-g^-$)-MHV amplitude $A(1,...,i,...,j,...,N|\mathrm{H})$ in EYM as:
\begin{equation}
\label{2.1}
A(1,...,i,...,j,...,N|\mathrm{H})\sim\frac{\langle ij\rangle^{4}}{\langle12\rangle\langle23\rangle\cdots\langle N1\rangle}S(i,j,\mathrm{H},\{1,...,N\})
\end{equation}

\noindent where $1$, ..., $N$ are gluons arranged in a fixed ordering, $\mathrm{H}=\{n_1,...,n_M\}$ are gravitons which are independent of color orderings. The negative-helicity gluons are supposed to be $i$ and $j$. The $S(i,j,\mathrm{H},\{1,...,N\})$ factor is generated by an exponential generating function, particularly
\begin{equation}
\label{2.4}
	\begin{aligned}
	S(\mathrm{H};\{1,...,N\})&=\qty(\prod\limits_{m\in\mathrm{H}}\frac{d}{da_{m}})\exp[\sum\limits_{n_{1}\in\mathrm{H}}a_{n_1}\sum\limits_{l\in \mathrm{G}}\psi_{ln_1} \exp[\sum\limits_{n_{2}\in\mathrm{H},n_{2}\neq n_{1}}a_{n_2}\psi_{n_1n_2}\exp(...)]]\Bigg|_{a_m=0},
	\end{aligned}
\end{equation}
in which 
\begin{equation}
\psi_{ab}\equiv\frac{[ab]\langle a\xi \rangle\langle a\eta \rangle}{\langle ab\rangle\langle b\xi \rangle\langle b\eta \rangle}\label{psi}
\end{equation}
where $\xi$, $\eta$ are arbitrarily chosen reference spinors and $\mathrm{G}$ is the gluon set. {\it In this note, we set $\xi=1$ and $\eta=N$ when studying the ($g^-,g^-$) configuration.}

It was shown in \cite{Du:2016wkt} that $S(\mathrm{H};\mathrm{G})$ could be expanded by spanning forest form. Particularly:
\begin{equation}
\label{2.5}
S(\mathrm{H};\mathrm{G})=\sum\limits_{F\in \mathcal{F}_{\mathrm{G}}(\mathrm{G}\cup\mathrm{H})}\qty(\prod\limits_{ab\in E(F)}\psi_{ab}),
\end{equation}
where we have summed over all possible forests $F$, where gluons and gravitons are considered as vertices, and the gluons are considered as the root set.
Each edge $ab$ is dressed by $\psi_{ab}$, and multiply all such edges in a given forest $F$ together.

{\it In the case of ($h^-,g^-$),  the formulas (\ref{2.1}) (\ref{2.4}) and (\ref{2.5}) are slightly changed \cite{Bern:1999bx,Du:2016wkt,Tian:2021dzf} via (i). replacing $i,j$ in (\ref{2.1}) by the negative helicity graviton and the negative-helicity gluon, (ii). replacing the gravitons set $H$ in (\ref{2.4}) and (\ref{2.5}) by the positive-helicity graviton set $\mathrm{H}^+$, while the root set is still the gluon set. (ii). introducing an extra minus $(-1)$.}

\section{Amplitudes with three gravitons} \label{sec:Example}
In this section, we extend the study of one and two graviton single-trace MHV amplitudes \cite{Chen:2010sr}, where each gravition is presented as a pair of collinear gluons, to the cases with an arbitrary number of gravitons. We demonstrate this by the example with three gravitons in the current section, and then provide a general formula in the next section.


%
\begin{figure}
\centering
\includegraphics[width=0.7\textwidth]{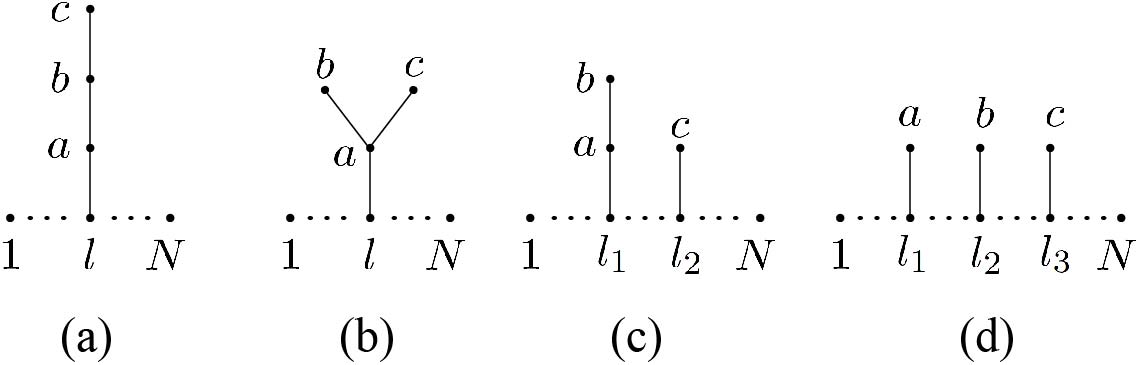}
\caption{All possible topologies of spanning forests for the three-graviton example. The $a$, $b$ and $c$ refer to different gravitons.}
\label{Fig:3GravitonTree}
\end{figure}
According to (\ref{2.1}) and (\ref{2.5}), the MHV amplitude with gluons $1$, ..., $N$ and three gravitons $n_1$, $n_2$, $n_3$ is  presented by  
\bea
A(1,...,N|n_1,n_2,n_3) \sim\frac{\langle ij\rangle^{4}}{\langle12\rangle\langle23\rangle\cdots\langle N1\rangle}\,S_3,
\eea
where $S_3$ is the abbreviation of the factor (\ref{2.5}) with three gravitons. Specifically, $S_3$ is expressed as
\begin{equation}
\label{2.7}
	\begin{aligned}
	S_3 = &\psi_1\psi_2\psi_3 + \psi_1\psi_2(\psi_{13}+\psi_{23}) + \psi_1\psi_3(\psi_{12}+\psi_{32}) + \psi_2\psi_3(\psi_{21}+\psi_{31}) \\
	&+ \psi_1(\psi_{12}\psi_{23}+\psi_{13}\psi_{32}+\psi_{12}\psi_{13}) + \psi_2(\psi_{21}\psi_{13}+\psi_{23}\psi_{31}+\psi_{21}\psi_{23})\\
	& + \psi_3(\psi_{31}\psi_{12}+\psi_{32}\psi_{21}+\psi_{31}\psi_{32}),
	\end{aligned}
\end{equation}
which are characterized by all possible spanning forests with structures \figref{Fig:3GravitonTree}. Each $\psi_{ab}$ ($a\neq b, a,b=1,2,3$) in the above expression is defined by (\ref{psi}) and is associating to an edge in the graphs \figref{Fig:3GravitonTree}, while the $\psi_i$ ($i=1,2,3$), associating to the graviton $n_i$, is defined by
\begin{equation}
\label{2.6}
 \psi_i\equiv\sum\limits_{l\in\mathrm{G}}\psi_{ln_i}.
\end{equation}
In the following, we analyze the contribution of each term in \eqref{2.7}.

{\it First}, let us deal with the term $\psi_1\psi_{12}\psi_{23}$, which is characterized by \figref{Fig:3GravitonTree} (a) (with $a=1$, $b=2$, $c=3$), on the right hand side of \eqref{2.7}. Noting that 
\begin{equation}
\label{2.8}
	\begin{aligned}
	\psi_1 = \sum\limits_{l\in \mathrm{G}} \frac{[ln_1]\langle l1\rangle\langle lN \rangle}{\langle ln_1 \rangle \langle n_11 \rangle \langle n_1N \rangle}&=\sum\limits_{l\in \mathrm{G}} [ln_1]\langle ln_1 \rangle \frac{-\langle 1l\rangle}{\langle 1n_1 \rangle \langle n_1l \rangle}\frac{\langle lN \rangle}{\langle ln_1 \rangle\langle n_1N \rangle}\\
	&=\sum\limits_{l\in \mathrm{G}} s_{ln_1} \sum\limits_{r_1=1}^{l-1}\frac{\langle r_1,r_1+1\rangle}{\langle r_1,n_1\rangle\langle n_1,r_1+1 \rangle}\sum\limits_{t_1=l}^{N-1}\frac{\langle t_1,t_1+1\rangle}{\langle t_1,n_1\rangle\langle n_1,t_1+1 \rangle},
	\end{aligned}
\end{equation}
where the eikonal identity (\ref{Eq:SpinorProp3}) and the fact that $s_{ln_1}=[ln_1]\langle n_1l \rangle $ are applied,
 we write the Parke-Taylor factor accompanied by $\psi_1\psi_{12}\psi_{23}$ as 
\bea
\label{2.9}
	\frac{\langle ij\rangle^{4}}{\langle12\rangle\langle23\rangle\cdots\langle N1\rangle}\psi_1\psi_{12}\psi_{23}
	%
	%
	%
	&=&\psi_{12}\psi_{23}\biggl[\,\sum\limits_{l\in \mathrm{G}}s_{ln_1}\sum\limits_{r_1=1}^{l-1}\sum\limits_{t_1=l}^{N-1}\langle ij\rangle^{4}\frac{1}{\langle12\rangle\cdots \langle r_1,n_1\rangle\langle n_1,r_1+1\rangle\cdots\langle l-1,l\rangle}\nn
	&&~~~~~~~~~~~~~~~~~~~~~~~~~~~~~~~\times \frac{1}{\langle l,l+1\rangle \cdots \langle t_1,\W n_1\rangle\langle \W n_1,t_1+1 \rangle \cdots \langle N1\rangle}\,\biggr],
\eea
where the factors $\langle r_1,r_1+1\rangle$ and $\langle t_1,t_1+1\rangle$ in the denominator of the Parke-Taylor factor have been replaced by $\langle r_1,n_1\rangle\langle n_1,r_1+1 \rangle$ and $\langle t_1,n_1\rangle\langle n_1,t_1+1 \rangle$, respectively. The $n_1$ in the second Parke-Taylor factor is further denoted by $\W n_1$. Hence, the graviton $n_1$ splits into two gluons $n_1$ and $\W n_1$ with the same momentum and helicity, which are respectively inserted between $1$, $l$ and $l$, $N$. Now we further express $\psi_{12}$ and  $\psi_{23}$ by 
\begin{equation}
\label{2.10} 
\psi_{12}=s_{n_1n_2}\sum\limits_{r_2=1}^{n_1-1}\frac{\langle r_2,r_2+1\rangle}{\langle r_2,n_2\rangle\langle n_2,r_2+1 \rangle}\sum\limits_{t_2=\W n_1}^{N-1}\frac{\langle t_2,t_2+1\rangle}{\langle t_2,n_2\rangle\langle n_2,t_2+1 \rangle},
\end{equation}
\begin{equation}
\label{2.11}
\psi_{23}=s_{n_2n_3}\sum\limits_{r_3=1}^{n_2-1}\frac{\langle r_3,r_3+1\rangle}{\langle r_3,n_3\rangle\langle n_3,r_3+1 \rangle}\sum\limits_{t_3=\W n_2}^{N-1}\frac{\langle t_3,t_3+1\rangle}{\langle t_3,n_3\rangle\langle n_3,t_3+1 \rangle},
\end{equation}
respectively.
\begin{figure}
\centering
\includegraphics[width=0.4\textwidth]{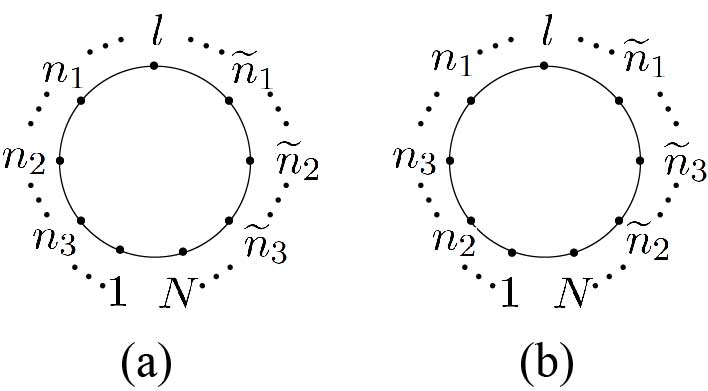}
\caption{(a). Permutations with the relative orderings $1,...,n_3,...,n_2,...,n_1,...,l,...,\W n_1,...,\W n_2,...,\W n_3,...,N$. (b). Permutations with the relative orderings $1,...,n_2,...,n_3,...,n_1,...,l,...,\W n_1,...,\W n_3,...,\W n_2,...,N$.}
\label{Fig:3Graviton1}
\end{figure}
When (\ref{2.10}) is substituted into (\ref{2.9}),  we find that the graviton $n_2$ splits into two gluons $n_2$ and $\W n_2$, which are respectively inserted to the left side of $n_1$ and to the right side of $\W n_1$. Similarly, (\ref{2.11}) finally inserts two gluons  $n_3$ and $\W n_3$ corresponding to the graviton $n_3$ to the left side of $n_2$ and the right side of $\W n_2$. The term $\frac{\langle ij\rangle^{4}}{\langle12\rangle\langle23\rangle\cdots\langle N1\rangle}\psi_1\psi_{12}\psi_{23}$ is then written as
\bea
\Sl_{l\in  \mathrm{G}}s_{n_1l}s_{n_2n_1}s_{n_3n_2}\Sl_{\pmb\rho^{(l)}}\text{PT}\left(1,\pmb\rho^{(l)},N\right),
\eea
in which, we introduced $\text{PT}\left(a_1,...,a_m\right)$ to denote the PT factor $\frac{\langle ij\rangle^{4}}{\langle a_1a_2\rangle\langle a_2a_3\rangle\cdots\langle a_ma_1\rangle}$ for short.  Permutations $\pmb\rho^{(l)}$ for a given $l\in\mathrm{G}$ are given by 
\bea
\pmb\rho^{(l)}\in\Big\{\{2,...,l-1\}\shuffle\{n_3,n_2,n_1\}, l, \{l+1,...,N-1\}\shuffle\{\W n_1,\W n_2,\W n_3\}\Big\},
\eea
which can be characterized by the graph \figref{Fig:3Graviton1} (a).

{\it Second}, we investigate the term with $\psi_1\psi_{12}\psi_{13}$, which is associated to the graph \figref{Fig:3GravitonTree} (b) (with $a=1$, $b=2$, $c=3$). When the factor $\psi_1$ and $\psi_{12}$ are expressed by (\ref{2.8}) and (\ref{2.10}), and $\psi_{13}$ is expressed as follows
\begin{equation}
\label{psi13} 
\psi_{13}=s_{n_1n_3}\sum\limits_{r_3=1}^{n_1-1}\frac{\langle r_3,r_3+1\rangle}{\langle r_3,n_3\rangle\langle n_3,r_3+1 \rangle}\sum\limits_{t_3=\W n_1}^{N-1}\frac{\langle t_3,t_3+1\rangle}{\langle t_3,n_3\rangle\langle n_3,t_3+1 \rangle},
\end{equation}
we just split the gravitons $n_1$, $n_2$ and $n_3$ into three pairs of gluons $\{n_1,\W n_1\}$, $\{n_2,\W n_2\}$ and $\{n_3,\W n_3\}$, respectively. The two gluons $n_1$, $\W n_1$ coming from the graviton $n_1$ are inserted to the left and the right sides of $l$, while the $n_2$ and $\W n_2$ (and also $n_3$ and $\W n_3$) are further inserted to the left of $n_1$ and the right of $\W n_1$. Thus this term turns into 
\bea
\frac{\langle ij\rangle^{4}}{\langle12\rangle\langle23\rangle\cdots\langle N1\rangle}\psi_1\psi_{12}\psi_{13}=\Sl_{l\in  \mathrm{G}}s_{n_1l}s_{n_2n_1}s_{n_3n_1}\Sl_{\pmb\rho^{(l)}}\text{PT}\left(1,\pmb\rho^{(l)},N\right),
\eea
where $\pmb\rho^{(l)}$ for a given $l$ is now given by 
\bea
\pmb\rho^{(l)}\in \Big\{\{2,...,l-1\}\shuffle\{\{n_3\}\shuffle\{n_2\},n_1\}, l, \{l+1,...,N-1\}\shuffle\{\W n_1,\{\W n_2\}\shuffle\{\W n_3\}\}\Big\},~~\Label{Eq:permsEG1}
\eea
which are characterized by \figref{Fig:3Graviton1} (a) and (b).

\begin{figure}
\centering
\includegraphics[width=0.75\textwidth]{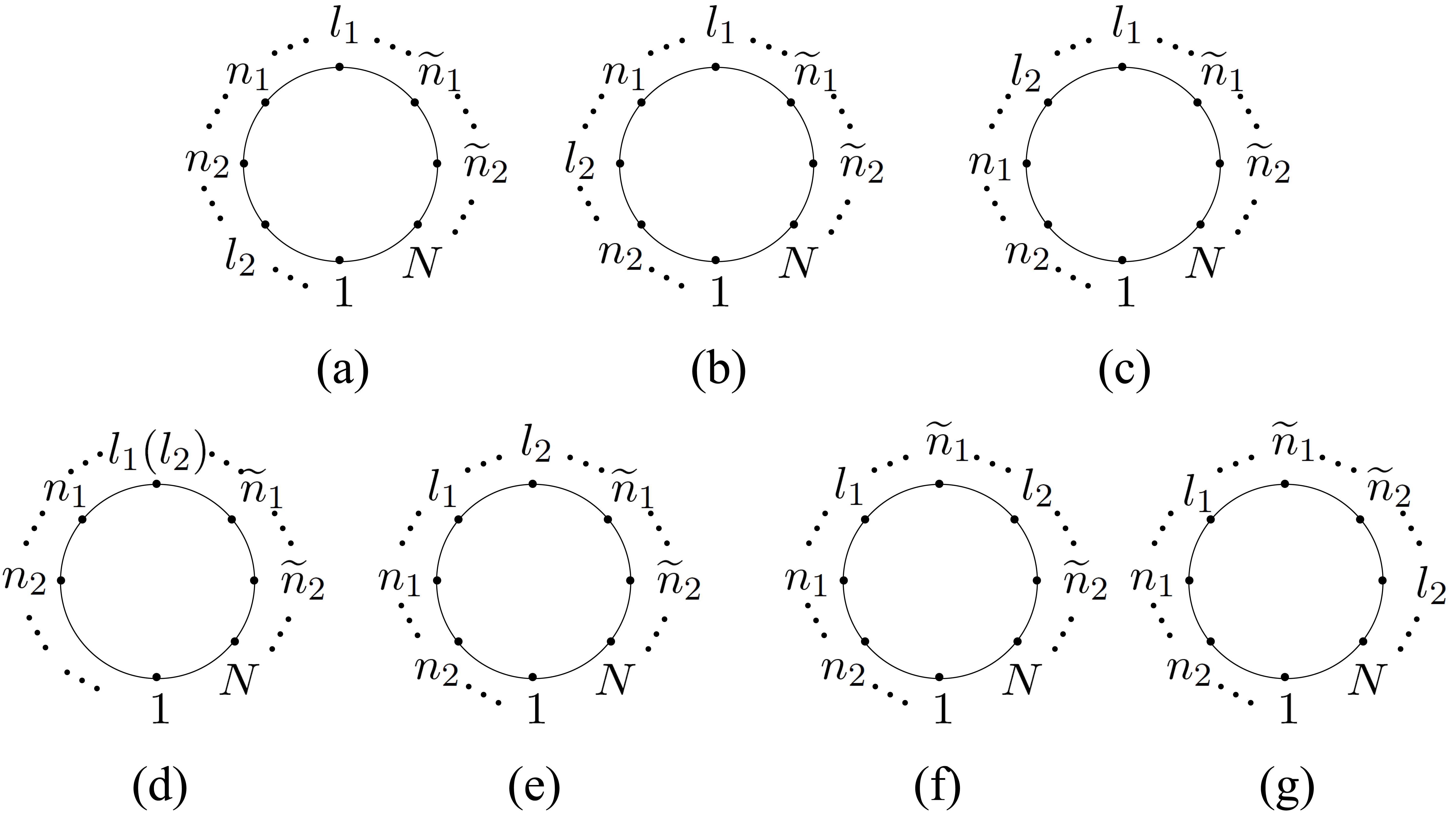}
\caption{Possible relative positions of $l_2$ in the permutations $\pmb\rho^{(l_1)}$ in (\ref{Eq:permsEG2})}
\label{Fig:3Graviton2}
\end{figure}
{\it Third}, we calculate the term with $\psi_1\psi_{3}\psi_{12}$ (see \figref{Fig:3GravitonTree} (c) with $a=1$, $b=2$, $c=3$). When the same trick with the previous examples is applied, $\psi_1$ and $\psi_{12}$ are expressed by (\ref{2.8}) and (\ref{2.10}), while $\psi_3$ is obtained via replacing $n_1$ in (\ref{2.8}) by $n_3$. Again, these factors are used to insert gluon pairs into the Parke-Taylor factor. The result is 
\bea
\frac{\langle ij\rangle^{4}}{\langle12\rangle\langle23\rangle\cdots\langle N1\rangle}\psi_1\psi_{3}\psi_{12}=\Sl_{l_1,l_2\in \mathrm{G}}s_{n_1l_1}s_{n_2n_1}s_{n_3l_2}\Sl_{\pmb\rho^{(l_1,l_2)}}\text{PT}\left(1,\pmb\rho^{(l_1,l_2)},N\right),
\eea
in which, $\pmb\rho^{(l_1,l_2)}$ for given $(l_1,l_2)$ satisfies
\bea
\pmb\rho^{(l_1,l_2)}&\in&\Big\{\pmb\rho^{(l_1)}_{L}\shuffle\{n_3\},l_2,\pmb\rho^{(l_1)}_R\shuffle\{\W n_3\}\Big\},~~\Label{Eq:permsEG2}\\
\text{where~}\pmb{\rho}^{(l_1)}&\in&\Big\{\{2,...,l_1-1\}\shuffle\{n_2,n_1\}, l_1, \{l_1+1,...,N-1\}\shuffle\{\W n_1,\W n_2\}\Big\}.\nonumber
\eea

%
On the second line, the $\pmb{\rho}^{(l_1)}$ denotes the permutations established by inserting the collinear gluons corresponding to $n_1$ and $n_2$ into the original gluon set, while $\pmb\rho^{(l_1)}_{L}$ and $\pmb\rho^{(l_1)}_{R}$ are the sectors separated by the gluon $l_2$ in the permutation $\pmb{\rho}^{(l_1)}$. Possible relative positions of $l_2$ in $\pmb{\rho}^{(l_1)}$ are displayed by \figref{Fig:3Graviton2} (a)-(g). Since the choices of $l_1$ and $l_2$ are independent of each other and we finally summed over all possible choices of $l_1$ and $l_2$, one can exchange the roles of $l_1$, $l_2$ in (\ref{Eq:permsEG2}) as follows
\bea
\pmb\rho^{(l_1,l_2)}&\in&\Big\{\pmb\rho^{(l_2)}_{L}\shuffle\{n_2,n_1\},l_1,\pmb\rho^{(l_2)}_R\shuffle\{\W n_1,\W n_2\}\Big\},~~\Label{Eq:permsEG2a}\\
\text{where~}\pmb{\rho}^{(l_2)}&\in&\Big\{\{2,...,l_2-1\}\shuffle\{n_3\}, l_2, \{l_2+1,...,N-1\}\shuffle\{\W n_3\}\Big\}.\nonumber
\eea

When all possible spanning forests for amplitude with three gravitons are considered, the full MHV amplitude with three gravitons is finally expressed by the following formula:
\bea
A(1,...,N|n_1,n_2,n_3)\sim \Sl_{\substack{\text{Spanning Forests}\\ \{\mathcal{T}_1,...,\mathcal{T}_i\}}}\,\Sl_{l_1,...,l_i\in\mathrm{G}}K(\mathcal{T}_1)\,...\,K(\mathcal{T}_i)\,\text{PT}\left(1,\pmb\rho^{(l_1,...,l_i)},N\right).~~~~(i\leq 3)\label{Eq:3GravitonSingleTrace}
\eea
In the above expression, we have summed over all possible spanning forests where the original gluon set $\mathrm{G}$ plays as the root set. For a given spanning forest with $i$ ($i\leq 3$) trees $\mathcal{T}_1,...,\mathcal{T}_i$ planted at gluons $l_1,...,l_i\in \mathrm{G}$ ($l_j$ and $l_k$ with distinct labels may be identical), each $K(\mathcal{T}_j)$ ($j=1,...,i$) is given by 
\bea
K(\mathcal{T}_j)=\prod\limits_{ab\in E(\mathcal{T}_j)}s_{ab},
\eea
where $ab\in E(\mathcal{T}_j)$ is an edge of the tree $\mathcal{T}_j$ with vertices $a$ and $b$.
More explicity, there are four possible topologies for the three-graviton amplitude, as shown by \figref{Fig:3GravitonTree} (a), (b), (c) and (d), which respectively provide factors 
\bea
s_{cb}s_{ba}s_{al},~~~~s_{ba}s_{ca}s_{al},~~~~s_{ba}s_{al_1}s_{cl_2},~~~~s_{al_1}s_{bl_2}s_{cl_3},
\eea
while  $a$, $b$, $c$ represent distinct gravitons. Two graphs with exchanging the branches attached to a same vertex are considered as the same graph,  e.g. \figref{Fig:3GravitonTree} (b). The permutations associated to \figref{Fig:3GravitonTree} (a) and (b) can be recursively defined by (\ref{Eq:permsEG1}) and (\ref{Eq:permsEG2}), via replacing the subscripts $1$, $2$ and $3$ of gravitons in (\ref{Eq:permsEG2}) by $a$, $b$ and $c$, respectively. The permutations for \figref{Fig:3GravitonTree} (c) satisfy
\bea
&&\pmb\rho^{(l_1,l_2)}\in\Big\{\pmb\rho^{(l_1)}\shuffle\{n_c\},l_2,\pmb\rho^{(l_1)}\cup\{\W n_c\}\Big\},\nn
&\text{where}&~\pmb\rho^{(l_1)}\in \Big\{\{2,...,l_1-1\}\cup\{n_b,n_a\},l_1,\{l_1+1,...,N-1\}\shuffle\{\W n_a,\W n_b\}\Big\}.
\eea
Permutations accompanying to \figref{Fig:3GravitonTree} (d) are presented by 
\bea
&&\pmb\rho^{(l_1,l_2,l_3)}\in\Big\{\pmb\rho^{(l_1,l_2)}\shuffle\{n_c\},l_3,\pmb\rho^{(l_1,l_2)}\shuffle\{\W n_c\}\Big\},\nn
&\text{where}&~\pmb\rho^{(l_1,l_2)}\in\Big\{\pmb\rho_L^{(l_1)}\shuffle\{n_b\},l_2,\pmb\rho_R^{(l_1)}\shuffle\{\W n_b\}\Big\}\nn
&\text{and}&~\pmb\rho^{(l_1)}\in\Big\{\{2,...,l_1-1\}\shuffle\{n_a\},l_1,\{l_1+1,...,N-1\}\shuffle\{\W n_a\}\Big\}.
\eea

Having displayed the example with three gravitons, we turn to the general formula in the next section.

\section{The general formula}\label{sec:GeneralFormula}
Inspired by the example in the previous section, we propose the following general formula where gravitions split into pairs of collinear gluons
\bea
A(1,...,N| \mathrm{H})\sim \Sl_{l_1,...,l_i\in\mathrm{G}}\,\Sl_{\substack{\text{Spanning Forests}\\ \{\mathcal{T}_1,...,\mathcal{T}_i\}}}\,K(\mathcal{T}_1)\,...\,K(\mathcal{T}_i)\,\text{PT}\left(1,\pmb\rho^{(l_1,...,l_i)},N\right).\label{Eq:GenSingleTrace}
\eea
Here we sum over all possible spanning forests in which trees are planted at gluons $l_1,...,l_i\in\mathrm{G}$. This summation is expressed by two summations:
\begin{itemize}
\item (i). summing over all possible choices of the roots $l_1,...,l_i$ ($i=1,...,M$), 
\item(ii). for a given choice of roots $l_1$ ,..., $l_i$, summing over all possible configurations of forests, which  consist of nontrivial trees  $\mathcal{T}_1$, ..., $\mathcal{T}_i$ planted at the gluons  $l_1$ ,..., $l_i$.
\end{itemize} 
For a fixed forest, each tree $\mathcal{T}_k$ is associated with a factor $K(\mathcal{T}_k)$ where each edge between two vertices $a$, $b$ is assigned by a factor $s_{ab}$. The permutations $\pmb\rho^{(l_1,...,l_k)}$ in the PT factors can be defined recursively:
\bea
\pmb\rho^{(l_1,...,l_{k})}=\left\{\,\pmb\rho_L^{(l_1,...,l_{k-1})}\shuffle\pmb\sigma^{\mathcal{T}_k},\,l_k,\,\pmb\rho_R^{(l_1,...,l_{k-1})}\shuffle \big(\W{\pmb\sigma}^{\mathcal{T}_k}\big)^T \right\}.~~ (k\leq i)\Label{Eq:GenPermutations}
\eea
where $\pmb\rho_L^{(l_1,...,l_{k-1})}$ and $\pmb\rho_R^{(l_1,...,l_{k-1})}$ denote the two ordered sets which are separated by the gluon $l_k$ in the permutation $\pmb\rho^{(l_1,...,l_{k-1})}$. The $\pmb\sigma^{\mathcal{T}_k}$ ($\W{\pmb\sigma}^{\mathcal{T}_k}$) stands for the permutations established by the tree graph $\mathcal{T}_k$ whose nodes are $\{n_i\}$ ($\{\W n_i\}$), while $(\W{\pmb\sigma}^{\mathcal{T}_k})^T$ denotes the reverse of $\W{\pmb\sigma}^{\mathcal{T}_k}$.

Now we sketch the proof of the general formula (\ref{Eq:GenSingleTrace}):
\begin{itemize}
\item (i). {\it Step-1} Expand the MHV amplitude according to (\ref{2.1}) and (\ref{2.5}) in terms of spanning forests. Each forest $F$ in general consists of $i$ tree structures $\mathcal{T}_1$, ..., $\mathcal{T}_i$ planted to gluons $l_1,...,l_i\in \mathrm{G}$.

\item (ii). {\it Step-2} For a given forest $F=\{\mathcal{T}_1$, ..., $\mathcal{T}_i\}$ and the tree $\mathcal{T}_1$, there are two types of edges (a). the edge between a graviton $a$ and the root (a gluon $l_1\in \mathrm{G}$), (b). The edge between two gravitons $b$ and $c$. In the former case, the edge is associated with a factor $\psi_{a}$ which is expressed according to (\ref{2.8}), while an edge of the latter form is accompanied by a factor $\psi_{bc}$, which is further rewritten as (\ref{2.10}). After this manipulation, the factor $\psi_{a}$ splits the graviton $n_a$ into collinear gluons $n_a$ and $\W n_a$ and then inserts them to the left and right of $l_1$, respectively. A factor $\psi_{bc}$ splits the graviton $n_c$ into collinear gluons $n_c$ and $\W n_c$ which are further inserted to the left of $n_b$ and the right of $\W n_b$ ($n_b$ which is nearer to root than $n_c$ has already been treated before). The factor assigned to each edge $bc$ is $s_{bc}$, and the product of all these factors gives $K(\mathcal{T}_1)$. The permutations established by this step are given by 
\bea
\pmb\rho^{(l_1)}=\left\{\{2,...,l_{1}-1\}\shuffle \pmb{\sigma}^{\mathcal{T}_1},l_1,\{l_1+1,...,N-1\}\shuffle \big(\W{\pmb{\sigma}}^{\mathcal{T}_1}\big)^T\right\}.
\eea

\item (iii). {\it Step-3} Insert the collinear gluons corresponding to the gravitons on trees $\mathcal{T}_2$, ..., $\mathcal{T}_i$ in turn, by repeating step-2.  We finally get the general formula (\ref{Eq:GenSingleTrace}) with permutations defined in (\ref{Eq:GenPermutations}).
\end{itemize}

\section{Conclusion}\label{sec:Conclusions}
In this note,  we presented a formula (\ref{Eq:GenSingleTrace}) for single-trace EYM amplitudes in the MHV configuration (with two negative-helicity gluons). Each graviton in this formula splits into a pair of collinear gluons. Thus an $N$-gluon, $M$-graviton amplitude is finally expressed as a combination of $N+2M$ gluon amplitudes with  $M$ pairs of collinear gluons. When the adjustment pointed in \secref{sec:Backgrounds} is considered, the formula (\ref{Eq:GenSingleTrace}) is straightforwardly extended to the MHV amplitude with one negative-helicity gluon and one negative-helicity graviton via (i). replacing $i$, $j$ in the numerator of the PT factor by the negative-helicity graviton and the negative-helicity gluon, (ii). using the positive-helicity graviton set instead of the full graviton set on the RHS of (\ref{Eq:GenSingleTrace}). (iii). dressing the expression by an extra sign $(-1)$.
It is worth extending the collinear expression in the current paper to the double-trace amplitudes and amplitudes with other helicity configurations in a future work.




\appendix

\bibliographystyle{JHEP}
\bibliography{reference.bib}

\end{document}